# Adaptive beamforming method based on recursive maximum correntropy in impulsive noise with alpha-stable process

Lu Lu • Haiquan Zhao

**Abstract-**As a well-established adaptation criterion, the maximum correntropy criterion (MCC) has been receiving increasing attention due to its robust against outliers. In this paper, a new complex recursive maximum correntropy (CRMC) algorithm without any priori information on the noise characteristics, is proposed under the MCC. The proposed algorithm is useful for adaptive beamforming, when the desired signal is contaminated by the impulsive noises. Moreover, the analysis of convergence property of the CRMC algorithm is performed. The results obtained from simulation study establish the effectiveness of this new beamformer.

## 1 Introduction

Adaptive beamforming has been widely applied in various wireless applications, for instance, radar [1], DS-CDMA system [2], and sonar [3]. It refers to simultaneously combine the signals from the elements of an array antenna, to suppress interference and provide target detection. Because of its low complexity, good convergence properties and satisfactory performance, the complex least mean squares (CLMS) algorithm has become one of the most popular adaptive beamforming algorithms. Furthermore, the application of the CLMS algorithm to the adaptive beamforming and its analysis have been extensively studied [4,5]. However, it may become unstable, when the desired signal is corrupted by impulsive noise. To surmount this shortcoming, several alternative methods were proposed for this case [6-8].

Many improved algorithms aimed at increasing the convergence speed of CLMS algorithm based mean square error (MSE) criterion have been presented in the literature. A constrained least mean-squares adaptive beamformer was proposed which utilizes a scheme to constrain the response of the beamformer undistorted [9]. In [10], Srar et. al proposed least mean square least mean square (LLMS) algorithm for adaptive array beamforming. Although this algorithm achieves an improved convergence performance over earlier CLMS-based algorithms, it has a relatively complex structure owing to using array image factor.

L. Lu • H. Zhao(✉)
School of Electrical Engineering, Southwest Jiaotong University, Chengdu, China.
e-mail: lulu@my.swjtu.edu.cn, hqzhao@home.swjtu.edu.cn

In practical environments, there are often impulsive noises, in which may have heavy tails and even may not possess finite second-order statistics. Based on [11], these noises can be modelled by $\alpha$-stable distribution in adaptive beamforming. In such case, the abovementioned MSE-based algorithms may fail to work. As a local similarity measure, the maximum correntropy criterion (MCC) has a close relationship with M-estimation, and insensitive to outliers [12]. It caught an increasing attention in recent years, due to its simplicity and robustness [13,14]. To deal with such an impulsive interference problem, the new complex recursive algorithm, called CRMC, is proposed based on MCC. Note that the proposed algorithm is particularly useful for adaptive beamforming, especially when the signals contain large outliers or contaminated by impulsive noises. The superior performance of CRMC is confirmed by simulation results about adaptive beamforming in $\alpha$-stable noise environments.

This paper is organized in the following manner. In section 2, a system model for adaptive beamforming in impulsive noise environment is presented. In section 3, we proposed the new CRMC algorithm. In section 4, the convergence analysis of proposed algorithm is performed. Section 5 presents some simulation results and Section 6 concludes the paper.

## 2 System model

Consider the spaced linear array of $M$ sensors receive a signal $\mathbf{x}(n)$ with known centre frequency of the narrowband signal $\omega$ and direction-of-arrival (DOA) $\theta_d$. The array measurement vector $\mathbf{x}(n)$ can be expressed as

$$\mathbf{x}(n) = \mathbf{a}(\theta_d)s_0(n) + \boldsymbol{\varepsilon}(n) \qquad (1)$$

where $\mathbf{a}(\theta_d) = [1, e^{j\omega\tau_1}, \ldots, e^{j\omega\tau_{M-1}}]^T$ is the steering vector, $\tau_i$ denotes time delay of the $i$th sensor relative to the first sensor, $j = \sqrt{-1}$ is the imaginary unit, $s_0(n)$ is the source signal snapshot, and the $M \times 1$ vector $\boldsymbol{\varepsilon}(n)$ is the additive noise vector.

When the noise is impulsive, it is often modeled as the symmetric $\alpha$-stable distribution which can be described by the characteristic function as follows [11]

$$\varphi(t) = \exp\{-|t|^\alpha\} \qquad (2)$$

where $0<\alpha<2$ is the characteristic exponent. The smaller $\alpha$ is, the more impulsive the process is. Assume that each element $\varepsilon(n) = \varepsilon_{Re} + j\varepsilon_{Im}$ of $\boldsymbol{\varepsilon}(n)$ follows an isotropic distribution [15] described by the following characteristic function:

$$E\{\exp(j\theta_1\varepsilon_{Re} + j\theta_2\varepsilon_{Im})\} = \exp\{-2^{-\alpha/2}(\sqrt{\theta_1^2 + \theta_2^2})^\alpha\} \qquad (3)$$

where $E\{\bullet\}$ stands for taking expectation.

## 3 Derivation of CRMC algorithm

Let $D$ and $Y$ be two random variables with the same dimensions, the measure of correntropy is defined as follows [12]

$$V(D,Y) = E\{\kappa(D,Y)\} = \int \kappa(d,y)\mathrm{d}\Re_{D,Y}(d,y) \qquad (4)$$

where $\kappa(\cdot,\cdot)$ is a shift-invariant *Mercer Kernel*, and $\Re_{D,Y}(d,y)$ denotes the joint distribution function of $(d, y)$. The most popular kernel used in correntropy is the Gaussian kernel:

$$\kappa(d,y) = \exp\left\{-\frac{|e|^2}{2\sigma^2}\right\} \tag{5}$$

where $e = d-y$, and $\sigma$ stands for the kernel size of correntropy. Here, we introduce $d$ to denote the desired signal, and $y$ denotes array outputs[1]. Then, the cost function of CRMC can be expressed as follows [12]

$$\begin{aligned}J_{CRMC}(n) &= \sum_{i=1}^{n}\lambda^{n-i}\kappa(d(i),y(i)) \\ &= \sum_{i=1}^{n}\lambda^{n-i}\exp\left\{-\frac{|e(i)|^2}{2\sigma^2}\right\}\end{aligned} \tag{6}$$

where $0 \ll \lambda < 1$ is the forgetting factor. Taking the gradient of $J_{CRMC}(n)$ with respect to the array coefficients $\mathbf{w}(n)$, we obtain

$$\frac{\partial J_{CRMC}(n)}{\partial \mathbf{w}(n)} = \sum_{i=1}^{n}\lambda^{n-i}\exp\left\{-\frac{|e(i)|^2}{2\sigma^2}\right\}e(i)\mathbf{x}(i). \tag{7}$$

Letting (7) to zero, one gets

$$\sum_{i=1}^{n}\lambda^{n-i}\psi(i)\mathbf{x}(i)\mathbf{x}^H(i)\mathbf{w}(n) = \sum_{i=1}^{n}\lambda^{n-i}\psi(i)\mathbf{x}(i)d(i) \tag{8}$$

where superscript $H$ denotes Hermitian operator (conjugate transpose). This differs from the standard solution for $l_2$ norm in the presence of weighting factors $\psi(i)$ given by

$$\psi(i) = \exp\left\{-\frac{|e(i)|^2}{2\sigma^2}\right\}. \tag{9}$$

Then, the expression of $\mathbf{w}(n)$ is obtained as follows:

$$\begin{aligned}\mathbf{w}(n) &= \mathbf{F}(n)\boldsymbol{\pi}(n) \\ &= \mathbf{R}^{-1}(n)\boldsymbol{\pi}(n)\end{aligned} \tag{10}$$

where $\mathbf{F}(n) = \mathbf{R}^{-1}(n)$, $\mathbf{R}(n) = \sum_{i=1}^{n}\lambda^{n-i}\psi(i)\mathbf{x}(i)\mathbf{x}^H(i)$ and $\boldsymbol{\pi}(n) = \sum_{i=1}^{n}\lambda^{n-i}\psi(i)\mathbf{x}(i)d(i)$. For $\psi(i) = 1$, the algorithm becomes the recursive least squares (RLS) algorithm. When $\psi(n) \neq 1$, $\mathbf{R}(n)$ and $\boldsymbol{\pi}(n)$ are themselves functions of the optimal weights via $\psi(n)$. It has to recalculate (10) in each iteration. To avoid this inconvenience, a sliding window method is proposed in [16]. However, the algorithm carries the main drawback of sliding-window strategy: the algorithm requires to keep in memory all previous samples within a window, making it is not a truly online algorithm. To further derive the truly online algorithm, $\mathbf{R}(n)$ is updated by recursive expression as follows:

$$\mathbf{R}(n) \approx \lambda\mathbf{R}(n-1) + \psi(n)\mathbf{x}(n)\mathbf{x}^H(n) \tag{11}$$

and

---

[1] Throughout this paper, $d(i)$ stands for desired signals $d$ at iteration $i$, $y(i)$ denotes array outputs $y$ at iteration $i$, respectively.

$$\pi(n) \approx \lambda\pi(n-1) + \psi(n)\mathbf{x}(n)d(n). \tag{12}$$

By using matrix inversion lemma [17], $\mathbf{F}(n)$ can be updated as

$$\mathbf{F}(n) = \lambda^{-1}\mathbf{F}(n-1) - \lambda^{-1}\boldsymbol{\Phi}(n)\mathbf{x}^H(n)\mathbf{F}(n-1) \tag{13}$$

where, the gain factor is defined by

$$\boldsymbol{\Phi}(n) = \frac{\psi(n)\mathbf{F}(n-1)\mathbf{x}(n)}{\lambda + \psi(n)\mathbf{x}^T(n)\mathbf{F}(n-1)\mathbf{x}(n)}. \tag{14}$$

From (10), (12), (13) and (14), $\mathbf{w}(n)$ can be updated as

$$\mathbf{w}(n) = \mathbf{w}(n-1) + \boldsymbol{\Phi}(n)\left[d(n) - \mathbf{x}^H(n)\mathbf{w}(n-1)\right]^*. \tag{15}$$

where * represents conjugate operation.

*Remark 1:* Note that (9), (14), and (15) define an implicit relationship between $\mathbf{w}(n)$ and $\psi(n)$ that cannot be solved in one step. Hence, the algorithm forces an iterative approximation to the solution, where $\psi(n)$ is calculated by using $\mathbf{w}(n-1)$, and the new value for $\mathbf{w}(n)$ is obtained via the value of $\psi(n)$.

*Remark 2*: The proposed algorithm is nearly blind since it does not require any priori information on the noise characteristics, and it can be implemented using only $\sigma$ and $\lambda$.

# 4 Analysis of the CRMC algorithm

## 4.1 Mean behavior analysis of the proposed algorithm

In this section, we perform the convergence analysis of the proposed algorithm based on MCC. First, the two assumptions are given as

*1) The desired response is produced by*

$$d(n) = \mathbf{w}_o^H\mathbf{x}(n) + e_\alpha(n). \tag{16}$$

Note that $\mathbf{w}_o$ is a vector containing the optimal coefficient values, and $e_\alpha(n)$ is the measurement noise. The noise signal $e_\alpha(n)$ and the array measurement vector $\mathbf{x}(n)$ are mutually independent.

*2) The array measurement vector $\mathbf{x}(n)$ is generated at random, the autocorrelation function is ergodic, which can be expressed as*

$$\boldsymbol{\Gamma} \approx \frac{1}{n}\mathbf{R}(n) \quad \text{if } n > M \tag{17}$$

where $\mathbf{R}(n)$ is a time average correlation matrix of $\mathbf{x}(n)$.

Assumption 2 implies that the time average can substitute for the ensemble average. Adding a regularization factor $\delta\lambda^n\mathbf{I}$ to $\mathbf{R}(n)$ yields:

$$\mathbf{R}(n) = \sum_{i=1}^{n}\lambda^{n-i}\psi(i)\mathbf{x}(i)\mathbf{x}^H(i) + \delta\lambda^n\mathbf{I}. \tag{18}$$

When $\lambda = 1$, we arrive at

$$\mathbf{R}(n) = \sum_{i=1}^{n}\psi(i)\mathbf{x}(i)\mathbf{x}^H(i) + \mathbf{R}(0). \tag{19}$$

Next, introducing (16) into (12), the vector $\boldsymbol{\pi}(n)$ can be written as

$$\boldsymbol{\pi}(n) = \sum_{i=1}^{n} \psi(i)\mathbf{x}(i)\mathbf{x}^H(i)\mathbf{w}_o(i) + \sum_{i=1}^{n} \psi(i)\mathbf{x}(i)e_\alpha^*(i). \tag{20}$$

Combining (19) and (20), we obtain

$$\boldsymbol{\pi}(n) = \mathbf{R}(n)\mathbf{w}_o - \mathbf{R}(0)\mathbf{w}_o + \sum_{i=1}^{n} \psi(i)\mathbf{x}(i)e_\alpha^*(i). \tag{21}$$

Hence, (19) is rewritten by

$$\begin{aligned}\mathbf{w}(n) &= \mathbf{R}^{-1}(n)\mathbf{R}(n)\mathbf{w}_o - \mathbf{R}^{-1}(n)\mathbf{R}(0)\mathbf{w}_o \\ &+ \mathbf{R}^{-1}(n)\sum_{i=1}^{n}\psi(i)\mathbf{x}(i)\mathbf{x}^H(i)e_\alpha^*(i).\end{aligned} \tag{22}$$

Taking the expectations of both sides of (22) and using assumptions 1 and 2, we obtain

$$\begin{aligned}E\{\mathbf{w}(n)\} &\approx \mathbf{w}_o - \frac{1}{n}\boldsymbol{\Gamma}^{-1}\mathbf{R}(0)\mathbf{w}_o \\ &= \mathbf{w}_o - \frac{\delta}{n}\boldsymbol{\Gamma}^{-1}\mathbf{w}_o \\ &= \mathbf{w}_o - \frac{\delta}{n}\mathbf{K} \quad (\text{if } n > M)\end{aligned} \tag{23}$$

where $\mathbf{K} = \boldsymbol{\Gamma}^{-1}\mathbf{w}_o$ is the average of cross-correlation vector between $d(n)$ and $\mathbf{x}(n)$. Consequently, according to the aforementioned analyses, the CRMC algorithm is convergent and stable. Due to using $\mathbf{F}(0) = (1/\sigma)\mathbf{I} = \delta\mathbf{I}$ to initialize the algorithm, the estimated value of $\mathbf{w}(n)$ is biased for $n > M$. However, when $n \gg M$ (i.e. $n \to \infty$), the estimator is unbiased and the estimation error tends to zero.

### 4.2 Mean weight behavior analysis of the proposed algorithm

To further verify the stability of the CRMC algorithm, the mean weight behavior analysis is conducted in this section. The weight deviation vector is defined as follows:

$$\boldsymbol{\Omega}(n) = \mathbf{w}_o - \mathbf{w}(n). \tag{24}$$

Combining (14) and (15), the tap-weight update formulation of the proposed algorithm can be rewritten as

$$\mathbf{w}(n) = \mathbf{w}(n-1) + \frac{\psi(n)\mathbf{F}(n-1)\mathbf{x}(n)}{\lambda + \psi(n)\mathbf{x}^H(n)\mathbf{F}(n-1)\mathbf{x}(n)}\left[d(n) - \mathbf{x}^H(n)\mathbf{w}(n-1)\right]. \tag{25}$$

Then, the update formulation of the weight deviation vector of the proposed algorithm can be expressed by using (24)

$$\boldsymbol{\Omega}(n) = \boldsymbol{\Omega}(n-1) - \frac{\psi(n)\mathbf{F}(n-1)\mathbf{x}(n)}{\lambda + \psi(n)\mathbf{x}^H(n)\mathbf{F}(n-1)\mathbf{x}(n)}\left[d(n) - \mathbf{x}^H(n)\mathbf{w}(n-1)\right]. \tag{26}$$

Taking expectation of both sides of (26), the mean convergence behavior of the coefficient vector can now be expressed by

$$E[\boldsymbol{\Omega}(n)] = E[\boldsymbol{\Omega}(n-1)] - E\left[\frac{\psi(n)\mathbf{F}(n-1)\mathbf{x}(n)}{\lambda + \psi(n)\mathbf{x}^H(n)\mathbf{F}(n-1)\mathbf{x}(n)}\left(d(n) - \mathbf{x}^H(n)\mathbf{w}(n-1)\right)\right]. \tag{27}$$

The priori error $\zeta(n) = d(n) - \mathbf{x}^H(n)\mathbf{w}(n-1)$ can be approximately calculated by

$$\zeta(n) \approx \mathbf{x}^H(n)\boldsymbol{\Omega}(n-1). \tag{28}$$

Introducing (28) to (27), and supposing the independence between the priori error $d(n) - \mathbf{x}^H(n)\mathbf{w}(n-1)$ and $\frac{\psi(n)\mathbf{F}(n-1)\mathbf{x}(n)}{\lambda + \psi(n)\mathbf{x}^H(n)\mathbf{F}(n-1)\mathbf{x}(n)}$, thus, (27) can be giving as

$$E[\mathbf{\Omega}(n)] = E[\mathbf{\Omega}(n-1)] - E\left[\frac{\psi(n)\mathbf{F}(n-1)\mathbf{x}(n)\mathbf{x}^H(n)}{\lambda + \psi(n)\mathbf{x}^H(n)\mathbf{F}(n-1)\mathbf{x}(n)}\mathbf{\Omega}(n-1)\right]$$
$$\approx E[\mathbf{\Omega}(n-1)] - E\left[\frac{\psi(n)\mathbf{F}(n-1)\mathbf{x}(n)\mathbf{x}^H(n)}{\lambda + \psi(n)\mathbf{x}^H(n)\mathbf{F}(n-1)\mathbf{x}(n)}\right]E[\mathbf{\Omega}(n-1)]$$
(29)

where $E\left[\frac{\psi(n)\mathbf{F}(n-1)\mathbf{x}(n)\mathbf{x}^H(n)}{\lambda + \psi(n)\mathbf{x}^H(n)\mathbf{F}(n-1)\mathbf{x}(n)}\right] \approx 0$. Therefore, the weight vector in CRMC converges if and only if

$$0 < \lambda_{\max}\left\{E\left[\frac{\psi(n)\mathbf{F}(n-1)\mathbf{x}(n)\mathbf{x}^H(n)}{\lambda + \psi(n)\mathbf{x}^T(n)\mathbf{F}(n-1)\mathbf{x}(n)}\right]\right\} < 2 \quad (30)$$

where $\lambda_{\max}\{\cdot\}$ denotes the largest eigenvalue of a matrix. According to the fact that $\lambda_{\max}(\mathbf{AB}) < \text{Tr}(\mathbf{AB})$ in (30), we obtain

$$\lambda_{\max}\left\{E\left[\frac{\psi(n)\mathbf{F}(n-1)\mathbf{x}(n)\mathbf{x}^H(n)}{\lambda + \psi(n)\mathbf{x}^H(n)\mathbf{F}(n-1)\mathbf{x}(n)}\right]\right\}$$
$$< E\left[\frac{\text{Tr}(\mathbf{x}(n)\mathbf{\Lambda}(n)\mathbf{x}^H(n))}{\lambda + \psi(n)\mathbf{x}^H(n)\mathbf{F}(n-1)\mathbf{x}(n)}\right] < 1$$
(31)

where $\mathbf{\Lambda}(n) = \psi(n)\mathbf{F}(n-1)$. Therefore, the mean error weight vector of the proposed algorithm is convergent if the input signal is persistently exciting [18].

### 4.3 Computational complexity

**Table 1** Computational complexity of the CLMS, Constrained LMP, CMPN, RLS, and CRMC for adaptive beamforming.

| Algorithms | Mul. | Add. | Other operations | Computational time(sec) |
|---|---|---|---|---|
| CLMS | $2M+1$ | $2M$ | 0 | 0.000106 |
| Constrained LMP | $2M+2$ | $2M-2$ | $p-2$ $p$-power operation | 0.000180 |
| CMPN | $2M+3$ | $2M+1$ | 2 logarithmic operation | 0.000150 |
| RLS | $2M^2+4M$ | $2M^2+2M$ | 0 | 0.000276 |
| CRMC | $2M^2+4M+5$ | $2M^2+2M$ | 1 exponential operation | 0.001081 |

The computational complexity of the CRMC is compared with that of the CLMS [4], RLS, constrained LMP [11] and the continuous mixed $p$-norm (CMPN) [8] algorithms in terms of the total number of multiplications, additions, other operations and computational time per recursion. Table 1 shows the numerical complexity of the algorithms in the presence of $α=1.2$. The increase in the complexity of the proposed algorithm compared with that of the RLS algorithm is moderate, and still with an affordable computation time.

# 5 Simulation results

To evaluate the performance of the proposed algorithm, the simulations are performed in the adaptive beamforming. Here, the relative error $10\log\{\|\mathbf{w}_o - \mathbf{w}(n)\|^2 / \|\mathbf{w}_o\|^2\}$ is employed to quantify the performance [7], where $\|\cdot\|$ denotes the $l_2$ norm, and the optimum space-time filter. For the simulations, the following conditions are considered.

- The linear array is equally spaced by half-wavelength.
- The noise follows the isotropic stable distribution with $α=1.2$ or $α=1.4$.

**Example 1.**

In this example, a desired quadrature phase-shift keying (QPSK) arrives at an angle of $15°$, and the QPSK interference signal arrives at $7°$ and $23°$ with the same amplitude as the desired signal (8 degree difference). The second and third examples are carried out with 16 elements (*M*=16). All the beamformers are shown for the 1000th iteration. In the CLMS algorithm, $μ$=0.0003 is selected, $p$=1, $μ$=0.001 are fixed for constrained LMP, and $μ$=0.001 is selected for CMPN. The forgetting factors of the RLS and CRMC are selected as 0.99 to guarantee the fast and stable convergence, and the kernel size $σ$ is set at 8.

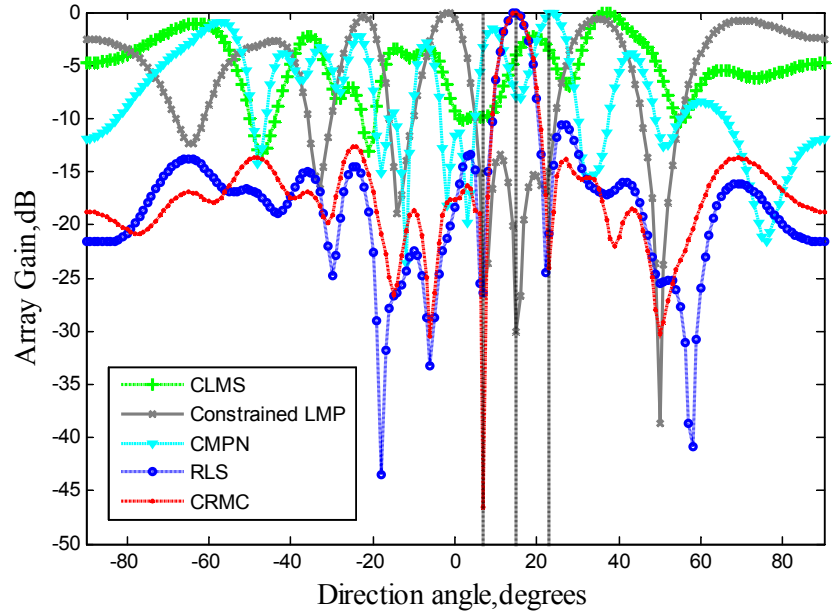

**Fig. 1** The beampatterns achieved with the CLMS, constrained LMP, CMPN, RLS and CRMC algorithms when the reference signal is contaminated by $α$-stable noise ($α$=1.2).

Fig. 1 illustrates the simulation results for beampatterns. It can be seen that the CLMS, constrained LMP and CMPN algorithms fail to work in this case, while the RLS and proposed algorithms have the stable performance. An important point in Fig. 1 need to be highlighted. Note from the QPSK interference signal at $7°$ that the suppression for CRMC algorithm is stronger than that of RLS algorithm. The validity of this confirms the fact that MCC estimation is much more robust against outliers than MSE estimation [12].

**Example 2.**

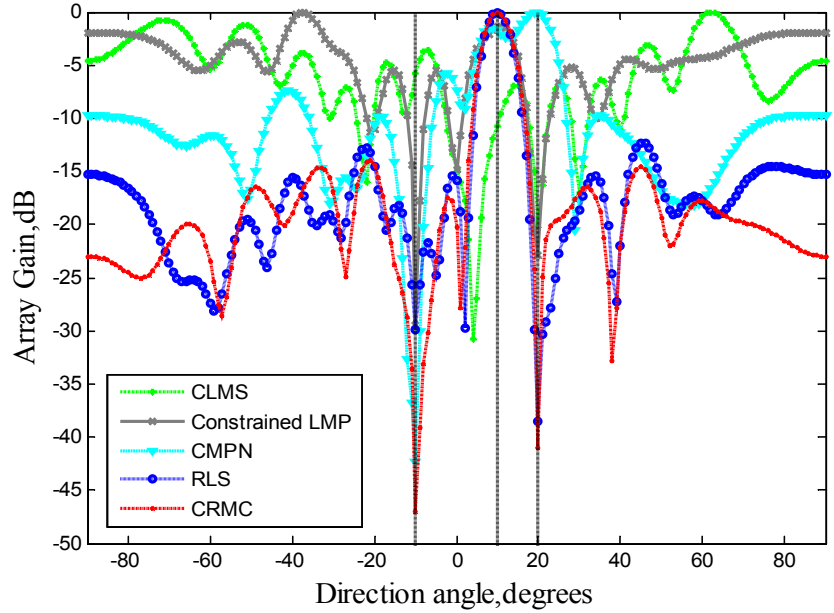

**Fig. 2** The beampatterns achieved with the CLMS, constrained LMP, CMPN, RLS and CRMC algorithms when the desired signal is $α$-stable noise ($α$=1.4).

We consider the case of desired signal with non-Gaussian distribution, and the reference signal is corrupted by $α$-stable noise. The proper selection of parameters for compared algorithms are the same as Example 1, the desired signal is $α$-stable noise ($α$=1.4), and the noise signal is $α$=1.2. The azimuth of the signal of interest is equal to $10°$ and the directions of the interfering signals are $-10°$ and $20°$. Fig. 2 illustrates the adaptive array beampatterns for algorithms. We see that the recursive-based algorithms, RLS and CRMC algorithms, outperform the other algorithms. Moreover, the CRMC beamformer achieves a strong null at the location of the interference ($-10°$ and $20°$).

From the simulation results of the above two scenarios, the proposed algorithm demonstrates the superior performance than the existing algorithms for both convergence rate and misadjustment. Also, correct beam patterns can still be achieved using the proposed algorithm. For highly impulsive noise process ($α$=1.2), the proposed algorithm enjoys better stability in comparison with LMS, constrained LMP and CMPN algorithms. For $α$=1.4, the CRMC algorithm provides small misadjustment and outperform the existing algorithms even for desired signal with non-Gaussian distribution.

## 6 Conclusion

Based on the MCC, a new CRMC algorithm, not requiring any *a priori* information, is proposed along with a Gaussian kernel for solving the adaptive beamforming problem. The MCC, which has been proven to be an efficient and robust optimization criterion for outliers, is used to improve the performance of beamformer. In addition, we study the mean behavior and mean weight behavior of the CRMC algorithm. As compared to the MSE-based criterion, the proposed algorithm achieves superior performance. Simulation results verified the efficiency of the proposed algorithm.

# Acknowledgments

This work was partially supported by National Science Foundation of P.R. China (Grant: 61571374, 61271340, 61433011).